\documentclass [aps, prb, reprint, floatfix, twocolumn,superscriptaddress]{revtex4}%
\usepackage{graphicx}
\usepackage{bm,amsmath,amssymb,mathrsfs,dcolumn}
\usepackage [colorlinks=true,linkcolor=blue,citecolor=blue,urlcolor=blue,linktoc=page,bookmarks=false,pdfstartview={FitH},pdfborder={0 0 0.0 [3 3]}]{hyperref}
\usepackage [usenames]{xcolor}
\hypersetup{pdfborder=0 0 0,colorlinks=true,citecolor=blue,linkcolor=blue}
\bibpunct{[}{]}{,}{n}{}{}

\def\be{\begin{equation}}
\def\ee{\end{equation}}
\def\ber{\begin{eqnarray}}
\def\eer{\end{eqnarray}}
\def\bs{\boldsymbol}
\begin{document}
\title{Dipolar spin waves in uniaxial easy-axis antiferromagnets: A natural topological nodal-line semimetal}
\author {Jie Liu}
\affiliation{The Center for Advanced Quantum Studies and Department of Physics, Beijing Normal University, Beijing 100875, China}
\author{Lin Wang}
\email{L.Wang-6@tudelft.nl}
\affiliation{Kavli Institute of Nanoscience, Delft University of Technology, P.O. Box 4056, 2600 GA Delft, The Netherlands}
\author {Ka Shen}
\email{kashen@bnu.edu.cn}
\affiliation{The Center for Advanced Quantum Studies and Department of Physics, Beijing Normal University, Beijing 100875, China}
\date{\today }


\begin{abstract}
  The existence of the magnetostatic surface spin waves in ferromagnets, known as Damon-Eshbach mode, was recently demonstrated to originate from the topology of the dipole-dipole interaction. In this work, we study the topological characteristics of magnons in easy-axis antiferromagnets with uniaxial anisotropy. The dipolar spin waves are found to be, driven by the dipole-dipole interaction, in a topological nodal-line semimetal phase, which hosts Damon-Eshbach-type surface modes due to the bulk-edge correspondence. The long wavelength character of dipolar spin waves makes our proposal valid for any natural uniaxial easy-axis antiferromagnet, and thus enriches the candidates of topological magnonic materials.  In contrast to the nonreciprocal property in ferromagnetic case, the surface modes with opposite momentum coexist at each surface, but with different chiralities. Such a chirality-momentum or spin-momentum locking, similar to that of electronic surface states in topological insulators, offers the opportunity to design novel chirality-based magnonic devices in antiferromagnets. 

\end{abstract}
\maketitle

{\it Introduction.---}Topological materials have attracted great interest in the past decade for their intriguing fundamental properties and promising applications. The related concepts, such as topological insulator~\cite{Hasan_RMP10,Qi_TI11} and Weyl semimetal~\cite{Vishwanath_RMP18}, in electronic materials have recently been introduced into magnonic systems~\cite{Shindou13,Shindou13B,Mook17,Jian18,Zhang:prb13,Mook16,SYing17,ChenGang18,SKKim16,SYing17b,Kovalev18,Pershoguba18,BYuan20,Mook14b,Chisnell15,Owerre18,FangChen17,Yuanli18,KSKim19}. Most of these proposals are based on quite selective materials with special crystal structures, such as pyrochlore~\cite{Zhang:prb13,Mook16,Mook17,SYing17,Jian18,ChenGang18}, honeycomb~\cite{SKKim16,SYing17b,Kovalev18,Pershoguba18,BYuan20} and Kagome lattices~\cite{Mook14b,Chisnell15,Owerre18}, and usually require Dzyaloshinskii-Moriya interaction~\cite{Zhang:prb13,Mook14b,Chisnell15,SKKim16,Mook16,FangChen17,SYing17,SYing17b,Owerre18,ChenGang18}, which supplies an effecitve spin-orbit interaction to magnons. While the topological magnons due to the dipole-dipole interaction, another interaction involving both spin and orbit degrees of freedom, have also been predicted in artificial magnonic crystals~\cite{Shindou13,Shindou13B}, the dipolar-induced topology in natural materials, especially for those without Dzyaloshinskii-Moriya interaction, is of particular interest and yet to be clarified.


As one of the most striking characteristics of the topological materials, topologically protected surface states appear at their boundaries. Among the dipolar spin waves in ferromagnets (FMs), on the other hand, there is also a special mode, called Damon-Eshbach mode~\cite{Damon1961}, with a very similar property of localization near the surface. This spin wave mode has recently attracted rising attention and is expected to be a useful ingredient for future magnonic devices~\cite{Jamali13,Mohseni19}. Although the Damon-Eshbach mode was discovered more than half a century ago, its topological origin was demonstrated until very recently by Yamamoto et al., who analyzed the bulk-edge correspondence from the topology of the dipolar interaction~\cite{Yamamoto19} and thus brought the concept of topology into the spin waves and their quanta, i.e., magnons, in {\it all} ferromagnetic materials.


For the magnons in FMs, however, only one spin state is allowed, which prevents the magnons to exhibit  desirable electron-like spin-orbit  phenomena. The latter perform an essential role in modern condensed matter physics. In contrast, antiferromagnets (AFMs) can host magnons polarized in two opposite directions as well as their superposition, and therefore activate the spin degree of freedom, suggesting that AFMs might be a suitable platform to study magnonic spin-orbit phenomena. The effective spin-orbit coupling due to the dipole-dipole interaction and its consequences, such as the D'yakonov-Perel'-type magnon spin relaxation mechanism and an intrinsic magnon (inverse) spin Hall effect, have already been predicted very recently~\cite{Shen20}.

In this paper, we show that the dipolar spin waves in uniaxial easy-axis AFMs are naturally in a topological nodal-line semimetal phase. More interestingly, the topological surface modes, residing in the dipolar-induced gap between the bulk bands, display a chirality-momentum locking, which is expected to be a useful property for applications.
Since the dipolar spin waves addressed here lie in the long wavelength regime and therefore are insensitive to the details of the crystal lattices, our results are valid for {\it any natural} easy-axis AFMs with uniaxial anisotropy. Based on the outstanding properties of the surface modes,  we propose a magnonic device to manipulate their chiralities through a magnetic field gradient.

{\it Bulk magnons in a uniaxial easy-axis AFM.---}In the presence of an external magnetic field along the easy axis, the magnons in an AFM with uniaxial anisotropy can be described by an effective two-band model under circularly polarized basis or spin basis, $\alpha$ and $\beta$, as~\cite{Shen20}
\be
H^{\rm eff}({\boldsymbol k})=\hbar\tilde\omega_{\boldsymbol k} +(\hbar\omega_H\hat z+\boldsymbol \Delta_{\boldsymbol k}^{\rm soc} )\cdot \boldsymbol \sigma,
\label{HH0}
\ee
with $\hbar\omega_H=\gamma\mu_0 H$ and $\boldsymbol \Delta_{\boldsymbol k}^{\rm soc}$ representing the Zeeman energy due to the external field and the effective magnon spin-orbit coupling from the dipolar interaction. As the precise dipolar field can be calculated from lattice  model (see Appendix~\ref{App_A}), in the long wavelength limit, it can be well described by $\boldsymbol \Delta_{\boldsymbol k}^{\rm soc}=\Delta_{\boldsymbol k}\sin^2\theta_{\boldsymbol k}(\cos{2\phi_{\boldsymbol k}},\sin{2\phi_{\boldsymbol k}},0)$~\cite{Shen20} with $\Delta_{\boldsymbol k}=\hbar\gamma_{\boldsymbol k}\omega_{\rm an}\omega_m/(2\tilde \omega_{\boldsymbol k})$ and
\be
\tilde \omega_{\boldsymbol k}=\sqrt{\omega_0^2+2\omega_{\rm an}A_{\boldsymbol k}+(1-\gamma_{\boldsymbol k}^2)(\omega_{\rm ex}+A_{\boldsymbol k})^2}.
\ee
Here, $\omega_0=\sqrt{\omega_{\rm an}(\omega_{\rm an}+2\omega_{\rm ex})}$ and $A_{\boldsymbol{k}}=(\omega_m/2)\sin^2 \theta_{\boldsymbol k}$ with $\omega_{\rm ex}$, $\omega_{\rm an}$ and $\omega_m$ representing the frequency scales of the exchange, anisotropy, and dipolar interaction, respectively. $\theta_{\boldsymbol k}$  corresponds to the polar angle of the wave vector ${\boldsymbol k}$ with respect to the easy axis along $z$ direction and  $\phi_{\boldsymbol k}$ stands for the azimuthal angle in the $x$-$y$ plane. The form factor $\gamma_{\boldsymbol k}=(1/{\cal Z})\sum_{\bs \delta}\exp(i \bs \delta \cdot \bs k)$ averages the phase factor over all coordination atoms. 

\begin{figure}[tp]
  \includegraphics[width=8.5cm]{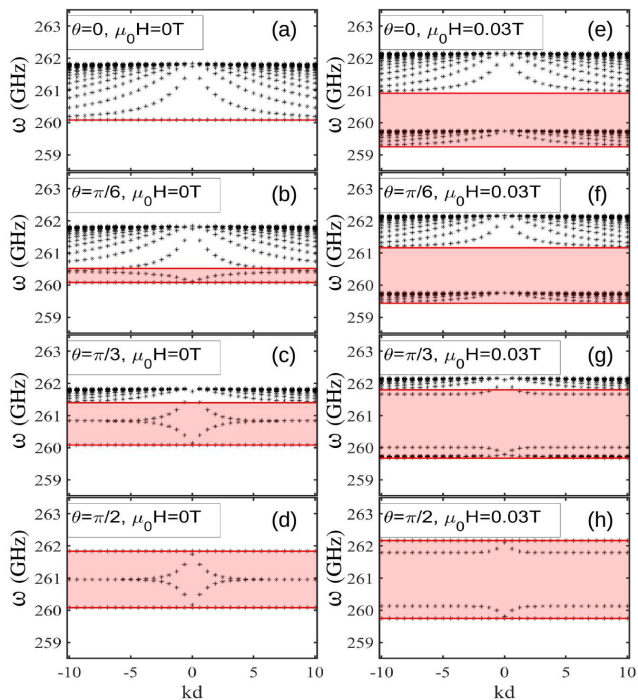}
  \caption{Spin wave spectra without (a-d) and with (e-h) external field for $\theta_{\boldsymbol k}=0$, $\pi/6$, $\pi/3$, and $\pi/2$. The red curves correspond to the frequency of the bulk modes given by Eq.~(\ref{bulkmode}). The parameters in MnF$_2$~\cite{Camley83} are adopted in the calculation with $\omega_{\rm ex}=1529$~GHz, $\omega_{\rm an}=21.96$~GHz, and $\omega_{m}=20.85$~GHz, giving $\omega_0\simeq 260.06$~GHz. And $\omega_H=0.836$~GHz for an external magnetic field $\mu_0H=0.03$~T.
  }
  \label{spectrum}
\end{figure}

Taking $\gamma_{\boldsymbol k}\sim 1$ near the zone center and assuming the dipolar interaction and the Zeeman term much smaller than $\omega_0$, the dispersion relations of the two magnon bulk bands can be calculated by diagonalizing Hamiltonian~(\ref{HH0}), leading to
\ber
(\omega_{\pm}^{\rm bulk})^2&=&\omega_{0}^{2}+\omega_{\rm an}\omega_{m}\sin^{2}\theta_{\boldsymbol k}\nonumber\\
&&\hspace{-1cm}\pm\sqrt{4\omega_{H}^{2}(\omega_{0}^{2} +\omega_{\rm an}\omega_{m}\sin^{2}\theta_{\boldsymbol k})+\omega_{\rm an}^{2}\omega_{m}^{2}\sin^{4}\theta_{\boldsymbol k}},
\label{bulkmode}
\eer
which are indicated by the red curves in Fig.~\ref{spectrum}. In the absence of external field, the bulk gap closes ($\omega_{+}^{\rm bulk}=\omega_{-}^{\rm bulk}$) at $\theta_{\boldsymbol k}=0$, forming a nodal line along $(0,0, k_z)$. Note that more nodal lines may appear in the Brillouin zone. For example, the lattice model in simple cubic lattice gives another nodal line along $(\pi/a_0,\pi/a_0,k_z)$  as shown in Appendix~\ref{App_B}. Although the specific crystal lattice may change the details of these additional nodal lines, it will not affect the long-wavelength dipolar spin waves  discussed in the present work.

{\it Topological analysis.---}By discarding the constant term $\tilde\omega_{\boldsymbol k}$ around ${\boldsymbol k}\simeq 0$~\cite{Yamamoto19}, the Hamiltonian (\ref{HH0}) reduces to  
\be
H^{\rm eff}({\boldsymbol k})\simeq \Delta_{\boldsymbol k}\sin^2\theta_{\boldsymbol k}(\cos2\phi_{\boldsymbol k} \sigma_x+\sin2\phi_{\boldsymbol k} \sigma_y),
\ee
for a vanishing magnetic field. This three-dimensional Hamiltonian has time-reversal symmetry (TRS) ${\boldsymbol T}H^{\rm eff}({\boldsymbol k}){\boldsymbol T}^{-1}=H^{\rm eff}({-\boldsymbol k})$, particle-hole like symmetry~\footnote{The particle-hole-like symmetry is specific for the two-band model, which is therefore different from the real particle-hole symmetry defined in the $4\times 4$ Hamiltonian in Appendix~\ref{App_A}.} ${\boldsymbol P}H^{\rm eff}({\boldsymbol k}){\boldsymbol P}^{-1}=-H^{\rm eff}({-\boldsymbol k})$, and therefore the chiral symmetry ${\boldsymbol C}H^{\rm eff}({\boldsymbol k}){\boldsymbol C}^{-1}=-H^{\rm eff}({\boldsymbol k})$. Here, ${\boldsymbol T}=\sigma_x{\boldsymbol K}$, ${\boldsymbol P}=\sigma_y{\boldsymbol K}$ and ${\boldsymbol C}=\sigma_z$ with ${\boldsymbol K}$ being the complex conjugation operator~\cite{Yamamoto19}. A gapped system of three dimensions satisfying such a symmetry property belongs to the topologically nontrivial class of CI~\cite{Schnyder08}. The present case is however gapless with nodal lines, therefore one has to reduce the three dimensional Brillouin zone to one-dimensional subsystems for topological classification~\cite{Sato11, Wang18}. The one-dimensional subsystem only has chiral symmetry in general~\cite{Schnyder08} and therefore is in class AIII, which can be characterized by the winding number $W(k_1, k_2)=({1}/{2\pi i})\int_{\rm (1D)BZ}{d\zeta (k_3)}/{\zeta(k_3)}$.
Here, $k_{1,2,3}$ are three orthogonal directions in momentum space and the integration for fixed $k_{1,2}$ is performed  over a closed loop formed by the one-dimensional Brillouin zone along $k_3$. Specifically for the present case, we have $\zeta(k_3)=e^{-2i\phi_{\boldsymbol k}}$ and therefore the reduced one-dimensional Hamiltonian is topological nontrivial unless $k_3$ is taken along the easy axis, in which $W(k_x, k_y)\equiv 0$ because of the fact that $\phi_{\boldsymbol k}$ is independent of $k_z$. As shown in Appendix~\ref{App_B}, the winding number can also be directly read from the variation of the spin-orbit field in momentum space, where its values around the two nodal lines $(0,0, k_z)$ and $(\pi/a_0,\pi/a_0,k_z)$ in simple cubic lattice are $-2$ and $2$, respectively.


In the following, we focus on a nontrivial configuration, in which the surfaces of  an AFM film is set to be normal to the $y$ axis.  Due to the chiral symmetry, all the surface modes have well defined chirality, equivalent to the winding number $W(k_x, k_z)$~\cite{Sato11}. For each surface, the two nodal lines  divide $W(k_x, k_z)$ into two blocks with either positive or negative $k_x$. These two blocks thus have opposite chiralities, i.e., $W(k_x, k_z)={\rm sgn}(k_x)$, indicating a chirality-momentum locking, similar to the spin-momentum locking in topological insulators~\cite{Hasan_RMP10,Qi_TI11}. This is because the TRS operator ${\boldsymbol T}$ anticommutes with chiral operator ${\boldsymbol C}$, i.e., $\{{\boldsymbol T}, {\boldsymbol C}\}=0$. In addition, for fixed momentum $(k_x, k_z)$, the chiralities of the modes at two surfaces are also opposite. This results from $\{{\boldsymbol M}_{xz}, {\boldsymbol C}\}=0$ with ${\boldsymbol M}_{xz}$ being the mirror symmetry about $xz$ plane ${\boldsymbol M}_{xz}H^{\rm eff}(y, k_x, k_z){\boldsymbol M}_{xz}^{-1}=-H^{\rm eff}(-y, k_x, k_z)$. Similar analysis can be done for other configurations (see Appendix~\ref{App_C} for the trivial case with surfaces normal to the easy axis).

With the external field included, the bulk Hamiltonian $H^{\rm eff}({\boldsymbol k})$ is gapped by the Zeeman term $\hbar\omega_H\sigma_z$ and only the particle-hole-like symmetry ${\boldsymbol P}$ survives. As a result, $H^{\rm eff}({\boldsymbol k})$ reduces to class C~\cite{Schnyder08}, which is topologically trivial in three dimensions.  The influence  of the external field on surface modes is rather interesting. The surface modes are eigenstates with respect to the external field and their frequencies are shifted up (down) with positive (negative) chirality.

{\it Spin wave spectrum in AFM film.---}Now we move to the detailed properties of the spin waves, of which the spectrum and wave functions can be derived by extending Damon-Eshbach approach~\cite{Damon1961} to AFM. Focusing on the dipolar spin wave regime~\cite{Camley80,Camley83,Camley10,Luthi83,Pereira99,AYu19}, the secular equation reads~\cite{Camley80,Camley83,Camley10}
\be [(1+\kappa)\frac {k_{y}^{i}}{|k|}\cot\frac{k_{y}^{i}d}{2}+1][(1+\kappa)\frac{k_{y}^{i}}{|k|}\tan\frac{k_{y}^{i}d}{2}-1]-\nu^{2}\sin^{2}\theta_{\boldsymbol k}=0, \label{Det_Freq}
\ee
where
\ber
(k_{y}^{i})^2&=&-\frac{1+\kappa\sin^{2}\theta_{\boldsymbol k}}{1+\kappa}k^{2},\\
\kappa&=&\frac{2\omega_{m}\omega_{\rm an}(-\omega_{H}^{2}+\omega_{0}^{2}-\omega^{2})}{[(\omega_{H}+\omega_{0})^{2}-\omega^{2}][(\omega_{H}-\omega_{0})^{2}-\omega^{2}]},\\
\nu&=&\frac{i4\omega\omega_{H}\omega_{m}\omega_{\rm an}}{[(\omega_{H}+\omega_{0})^{2}-\omega^{2}][(\omega_{H}-\omega_{0})^{2}-\omega^{2}]},
\eer
with $d$ being the thickness of the film.  The condition for the magnetostatic surface spin waves (MSSW) is then $(k_y^i)^2<0$, corresponding to a pure imaginary wave vector along the thickness direction.

The entire spin wave spectrum for an arbitrary $\theta_{\boldsymbol k}$ can be calculated numerically. The results for several typical angles without and with external field are plotted in Fig.~\ref{spectrum}. The surface modes residing in the magnon gap (represented by the red shadow) are clearly seen, except for $\theta_{\boldsymbol k}=0$ along the nodal line. At vanishing field, the two surface modes become degenerate at $|k|d\gg 1$ with the frequency
\be
(\omega_{\pm}^{\rm MSSW})^2|_{|k|d\gg 1}=\omega_{0}^{2}+2\omega_{\rm an}\omega_{m}\frac{\sin^{2}\theta_{\boldsymbol k}}{1+\sin^{2}\theta_{\boldsymbol k}},
\label{omega_theta}
\ee
which explains well the  angular dependence shown in Fig.~\ref{spectrum}. The splitting between the two surface modes in the small $|k|d$ regime results from the interplay of the two surfaces separated by a finite thickness. Note that the dipolar interaction alone can only lift the degeneracy of the upper set of the volume modes, but not the lower set, which is of a fixed frequency $\omega_0$  for any $\theta_{\boldsymbol k}$ [see Fig.~\ref{spectrum}(a-d)]. The inclusion of an external field not only introduce splitting to the low set but also pushes them into the bulk gap [see Fig.~\ref{spectrum}(e-h)]. In the following, we focus on the configuration with a largest bulk gap at $\theta_{\boldsymbol k}=\pi/2$, commonly used in experiment on the Damon-Eshbach spin waves in FMs~\cite{Jamali13}. The two surface modes in this configuration can be calculated analytically.

{\it Chiral surface spin waves at $\theta_{\boldsymbol k}=\pi/2$.---}For this particular case, we derive the solutions of the two surface modes from Eq.~(\ref{Det_Freq}) as
\ber
(\omega_{\pm}^{\rm MSSW})^2 &=&\omega_{H}^{2}+\omega_{0}^{2}+\omega_{\rm an}\omega_{m}\nonumber\\
&&\hspace{-0.7cm}{}\pm\sqrt{4\omega_{H}^{2}(\omega_{0}^{2}+\omega_{\rm an}\omega_{m})+\omega_{\rm an}^{2}\omega_{m}^{2}e^{-2|k|d}}, \label{omegapm}
\eer
which, in the limit of $|k|d\gg 1$, reduce to
\be
(\omega_{\pm}^{\rm MSSW})^2|_{|k|d\gg 1} =(\sqrt{\omega_{0}^2+\omega_{\rm an}\omega_m}\pm\omega_H)^2.
\ee
The two surface modes are only split by a Zeeman term. In the absence of external field, the two modes become degenerate, consistent with Eq.~(\ref{omega_theta}). In the opposite limit, $|k|d\ll 1$, Eq.~(\ref{omegapm}) become
\ber
(\omega_{\pm}^{\rm MSSW})^2|_{|k|d\ll 1} &=&\omega_{H}^{2}+\omega_{0}^{2}+\omega_{\rm an}\omega_{m}\nonumber\\
&&\hspace{-0.5cm}{}\pm\sqrt{4\omega_{H}^{2}(\omega_{0}^{2}+\omega_{\rm an}\omega_{m})+\omega_{\rm an}^{2}\omega_{m}^{2}}, 
\eer
which approach to the frequencies of the bulk modes determined by Eq.~(\ref{bulkmode}).

\begin{figure}[tp]
  \includegraphics[width=8cm]{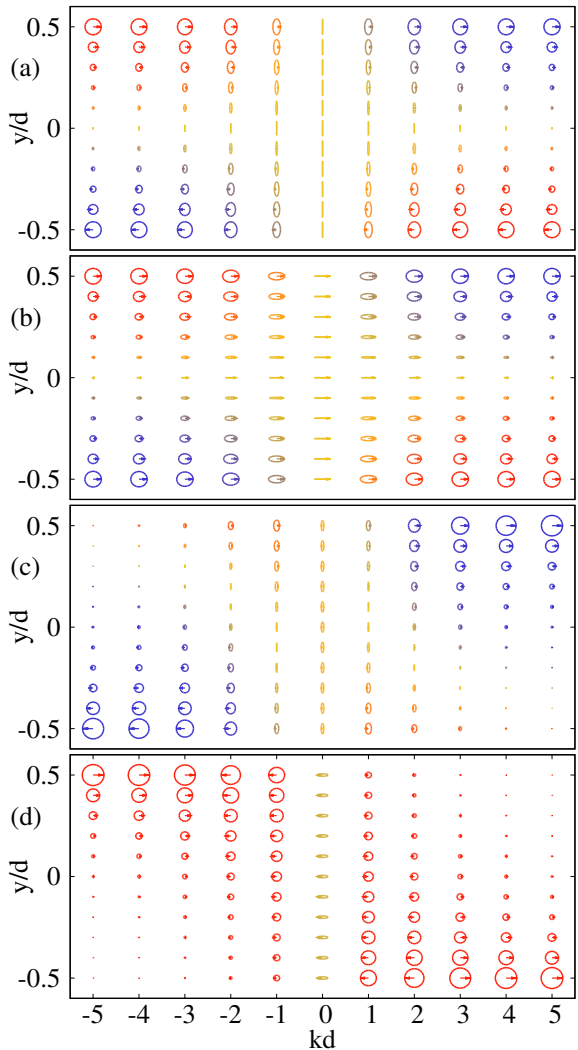}
  \caption{The spatial distributions of the magnetization dynamics for the surface modes at $\theta_{\boldsymbol k}=\pi/2$ (a, b) without and (c, d) with applied magnetic field ($H>0$). The blue and red colors characterize the local dynamics of the clockwise and counterclockwise chirality. (a, c) and (b, d) are for $\omega_+^{\rm MSSW}$ and $\omega_-^{\rm MSSW}$ modes, respectively. The arrows in each column show an instantaneous spin configuration of each mode. }
  \label{wave}
\end{figure}

The extension of Damon-Eshbach theory into AFM also gives the spatial profiles of the mode-dependent magnetization dynamics~\cite{Hashimoto17,Shen_2018JPD}. In Fig.~\ref{wave}, we plot the trajectories of the local magnetization dynamics across the film for the two surface modes with different values of $kd$. The size and color of the trajectories represent the amplitude and local chirality, respectively. The results without external field, i.e., Figs.~\ref{wave}(a) and (b) can be expressed generally as a combination of two chiral motions~\cite{Shen18}
\be
(m_{x},m_{y})_{\pm}^{\rm MSSW}(y)=e^{ky}(1,-i)\mp e^{-ky}(1,i),
\ee
with the relative magnitude of the $(1,-i)$ (clockwise motion) and $(1,i)$ (counterclockwise motion) components weighted by the colors, blue and red, in the figure. As seen from Figs.~\ref{wave}(a) and (b), the top (bottom) surface is dominated by the clockwise (counterclockwise) motion for a positive $k$. The rotation direction is reversed when the wave vector changes to the opposite direction. And the amplitudes at the top and bottom surfaces for each mode are always equal. These properties are consistent with our symmetry analysis. 

By introducing an external magnetic field to break the TRS, one can make each surface mode localize individually at only one surface as shown in Fig.~\ref{wave}(c) and (d). The situation becomes quite similar to the Damon-Eshbach modes in FMs, where the TRS is broken naturally. More interestingly, in the homogeneous limit, i.e., $k\simeq 0$, both surface modes become almost linearly polarized, which results from the superposition of the two dynamic components with comparable magnitudes.

\begin{figure}[tp]
  \includegraphics[width=9cm]{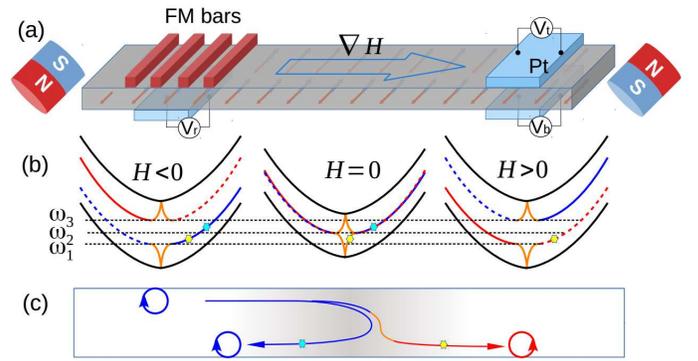}
  \caption{(a) Schematic of chirality inverter controlled by a magnetic field gradient. The periodically arranged FM bars form a momentum-selective spin wave generator and the Pt bars are detectors of the net local angular momentum. (b) Spin wave dispersion relations in different regimes. The solid and dashed curves represent the surface spin wave mode localized at top and bottom surfaces, respectively. The red and blue colors are to distinguish the chirality. (c) The trajectory of the surface spin wave propagation.}
  \label{setup}
\end{figure}

{\it Devices design.---}The features discussed above offer the opportunity to introduce the chirality degree of freedom as an ingredient of future AFM magnonic devices. As an example of potential applications, we propose a chirality inverter with an in-plane easy-axis AFM film~\cite{JShi20,Vaidya20}, as illustrated in Fig.~\ref{setup}, where a magnetic field gradient is applied along the magnon propagating channel to establish a spatial evolution of the dispersion.  Such a spatial profile of the external field can be realized experimentally by the two oppositely oriented magnets at the two ends~\cite{Rezende18}. As shown by Fig.~\ref{setup}(b), the chirality of the lower surface branch at $|k|d\gg 1$ is reversed (from clockwise to counterclockwise) when the magnetic field varies  from the negative to positive. Note that the parabolic feature due to the exchange interaction, which is neglected in the above calculations, has been taken into account in the short-wavelength dispersion in Fig.~\ref{setup}(b).

We consider the spin wave propagation excited by the dipolar field of a FM nanowire array associated with a microwave~\cite{Chen19}. The wavelength and the frequency of the active spin wave should match the period of the nanowires and the selected frequency of the driving microwave, respectively~\cite{Chen19,Han2020}. For a driving microwave lying in the frequency window between $\omega_1$ and $\omega_2$, the excited spin wave labeled by the yellow dot in the left dispersion of Fig.~\ref{setup}(b) is assumed to present pure clockwise magnetization rotation localized near the top surface. As the spin wave propagates from left to right, owing to its positive group velocity, its wave vector is modified but its frequency keeps unchanged. In the area of nearly zero magnetic field, as shown by the yellow dot in the middle dispersion of Fig.~\ref{setup}(b), the spin wave adiabatically evolves to be a mixture of clockwise and counterclockwise rotations. This state is no longer restricted near the top surface but extends across the entire thickness, as shown in Fig.~\ref{wave}. When the spin wave propagates further to the positive field regime, it shrinks again to one of the surfaces, but to the bottom one instead of back to the top one. In the meantime, the rotation direction of the magnetization so as the chiral current is reversed, as shown in Fig.~\ref{setup}(c). In contrast, if the driving frequency is tuned to be between $\omega_2$ and $\omega_3$, the excited spin wave, labeled by the light blue dot in the dispersion figures, is not able to flip its polarization, therefore, cannot transmit to the positive magnetic field side because of the increasing potential barrier. As a result, the spin wave goes through the thickness to the bottom surface and reflects back to the left side, as illustrated in Fig.~\ref{setup}(c). Since the propagating direction is reversed while the polarization remains the same, the chiral current is also reversed.

To experimentally test our proposal, the inverse spin Hall effect of heavy metals, which can translate the magnon spin polarization into electric signal, could be a promising way~\cite{Shen2019c}. Specifically, the DC inverse spin Hall voltages $V_t$, $V_b$, and $V_r$ of the three Pt bars in Fig.~\ref{setup}(a) are expected to display distinguishable behaviors. For $\omega_1<\omega<\omega_2$, $V_b$ would be of much larger amplitude than $V_r$ and $V_t$. As the frequency increases to the window $\omega_2<\omega< \omega_3$, the reflection process becomes dominant, making $V_r$ of the largest magnitude. Besides, $V_r$ and $V_b$ should have opposite signs.


{\it Summary.---}We have analyzed the topological nature of magnons in an easy-axis antiferromagnet with uniaxial anisotropy, which is recognized as a topological nodal-line semimetal driven by the dipolar interaction. We find a spin-momentum locking of the surface modes, which not only reveals the topological protection of these modes, but also paves the way to design novel AFM-based devices. Since the chirality degree of freedom addressed can convert into spin polarization of mobile electrons in a neighboring metal, the physics discussed in the present work may play a notable role in the spin transport and dynamics in heterostructures consisting of antiferromagnetic insulators and heavy metals~\cite{JShi20,Vaidya20}. A proposal of experimental setup to generate, manipulate, and detect chiral current is addressed. Finally, we want to point out that the qualitative features of the dipolar spin waves discussed here is expect to be robust against the thermal fluctuation and the magnon-magnon interactions, just like the situation in ferromagnets.

\begin{acknowledgments}
  This work is supported by the Recruitment Program of Global Youth Experts, the National Natural Science Foundation of China (Grants No.11974047), the Fundamental Research Funds for the Central Universities (Grant No. 2018EYT02) and the Netherlands Organisation for Scientific Research (NWO/OCW), as part of 
the Frontiers of Nanoscience program.

\end{acknowledgments}

\appendix
\section{Calculation of dipolar-induced splitting from a lattice model} \label{App_A}
\begin{figure}[ptb]
  \includegraphics[width=5.5cm]{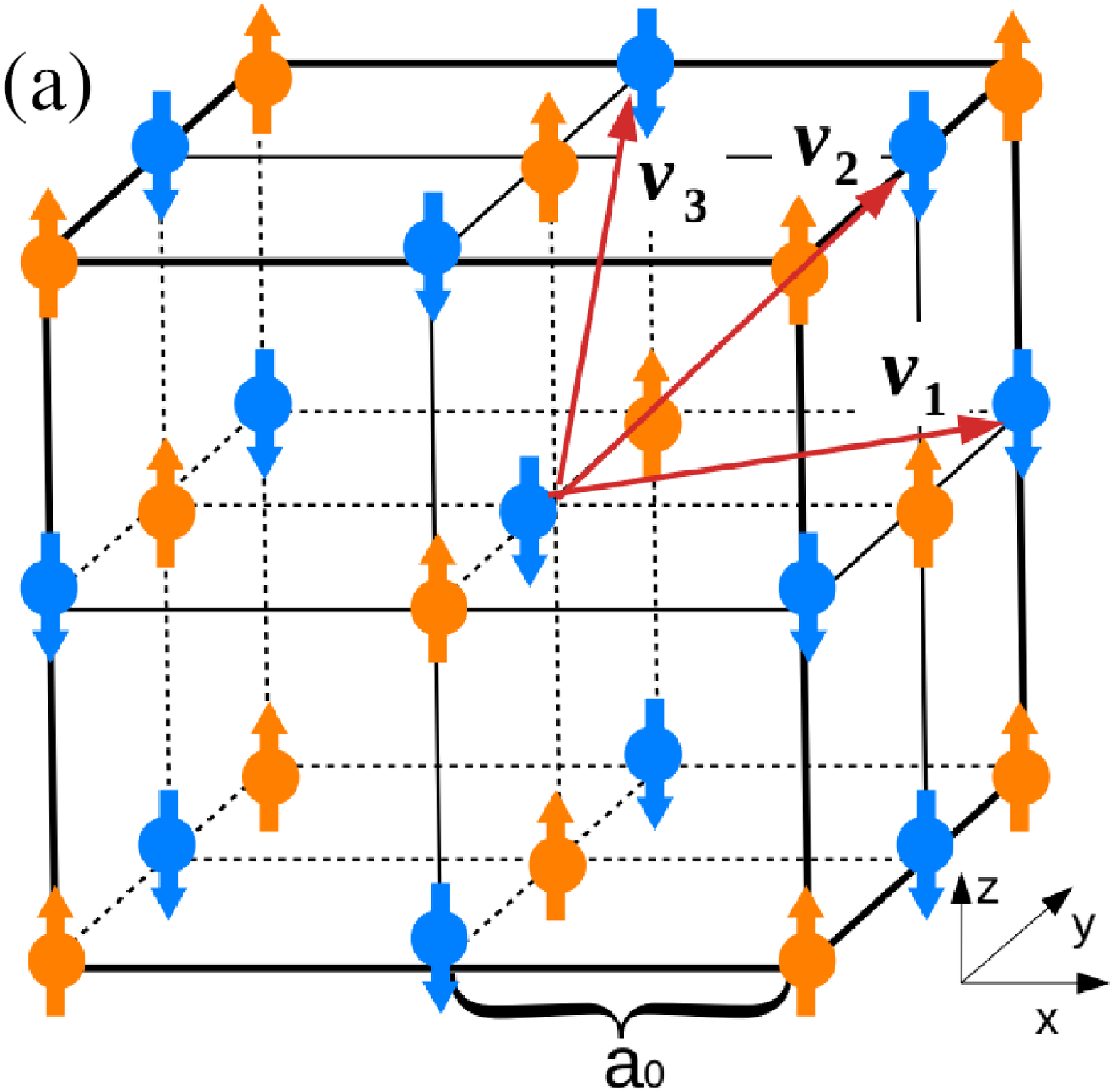}
  \includegraphics[width=8cm]{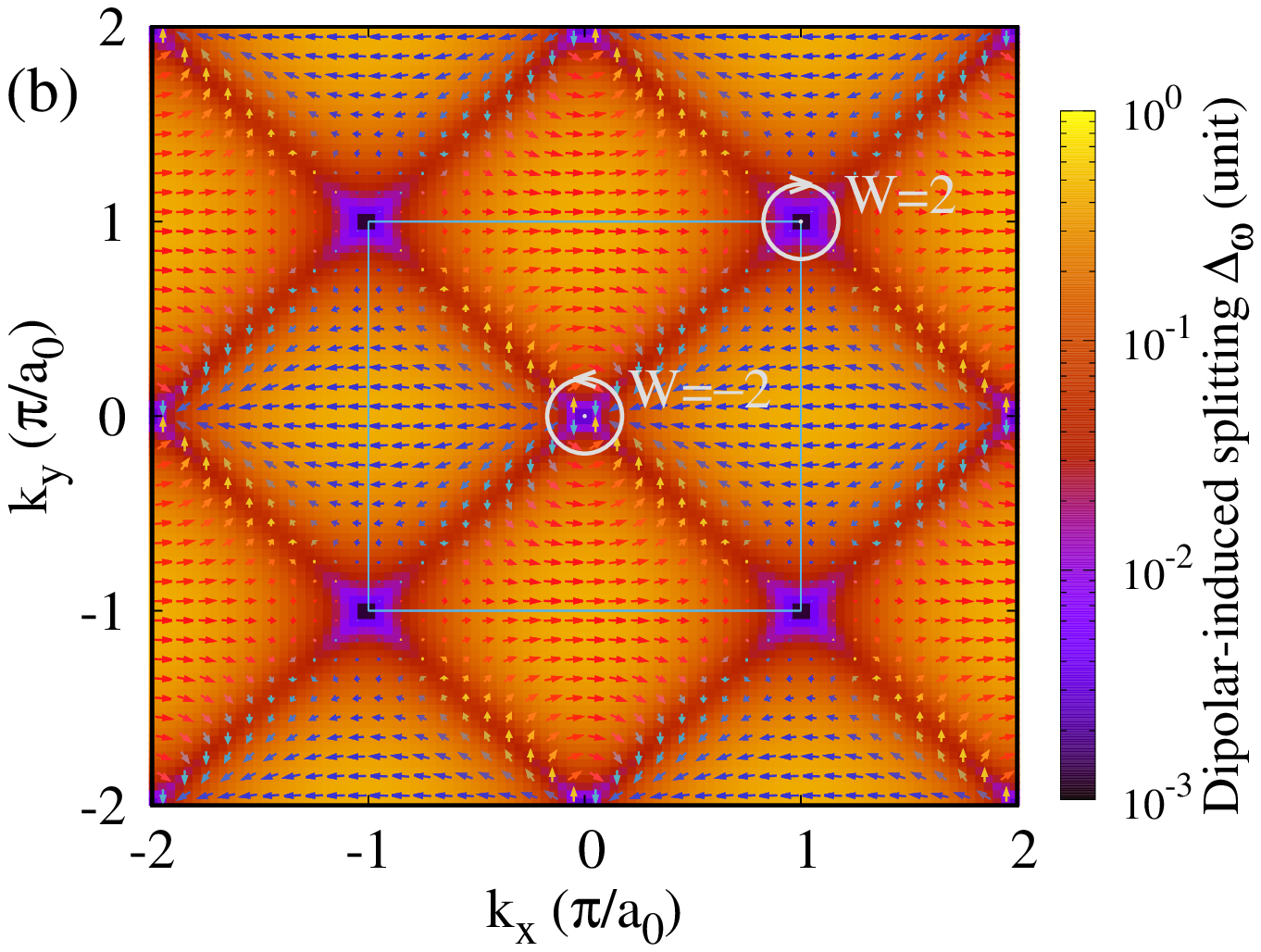}
  \caption{(a) Simple cubic lattice with all nearest neighboring magnetic atoms of opposite spin orientation. The translation vectors are defined as ${\boldsymbol v}_1=a_0{(1,1,0)}$, ${\boldsymbol v}_2=a_0{(1,0,1)}$, and ${\boldsymbol v}_3=a_0{(0,1,1)}$ with $a_0$ being the lattice constant. (b) The dipolar-induced splitting (background color) and its effective spin-orbit field (arrows) in momentum space in simple cubic lattice at $k_z= 0.2\pi/a_0$. The size of the arrows represents the strength of the field. The light blue square stands for the boundary of the first Brillouin zone. The phase change for the two opposite winding loops around two nodal lines $(0,0, k_z)$ and $(\pi/a_0,\pi/a_0,k_z)$ are both $4\pi$, indicating their  winding number $W=\mp 2$. }
  \label{Splitting}
\end{figure}

We consider a simple cubic lattice illustrated in Fig.~\ref{Splitting}, where all nearest neighboring spins are anti-parallel. The Hamiltonian of the dipole-dipole interaction reads
\be
H^{\rm DDI}=\frac{\mu_0 (g\mu_B)^2}{2}\sum_{l\ne l'} \frac{{|{\boldsymbol R}_{ll'}|}^2 \boldsymbol S_l\cdot\boldsymbol S_{l'}-3({{\boldsymbol R}_{ll'}}\cdot \boldsymbol S_l)({{\boldsymbol R}_{ll'}}\cdot \boldsymbol S_{l'})}{|{\boldsymbol R}_{ll'}|^5},\label{dd}
\ee
which includes contributions from any spin pairs.  By applying the Holstein-Primakoff transformation to the two spin sublattices
\ber
S_{a}^{z}=S-a^{\dagger}a,& S_{a}^{+}=\sqrt{2S-a^{\dagger}a}a,\nonumber\\
S_{d}^{z}=-S+d^{\dagger}d,& S_{d}^{+}=d^{\dagger}\sqrt{2S-d^{\dagger}d},
\eer
we can rewrite Eq.~(\ref{dd}) in the momentum space under the basis of $(a_{\boldsymbol k},d_{\boldsymbol k},a^\dagger_{-\boldsymbol k},d^\dagger_{-\boldsymbol k})^T$ as
\begin{equation}
  H_{\boldsymbol{k},-\boldsymbol{k}}=\left(\begin{array}{cccc}
{{ A}_{\mathbf{k}}^{aa}} & B_{-\boldsymbol{k}}^{ad\ast} & B_{\mathbf{k}}^{aa\ast} & A_{\boldsymbol{k}}^{ad\ast}\\
{B_{-\boldsymbol{k}}^{ad}} & {A}_{\mathbf{k}}^{dd} & A_{\boldsymbol{k}}^{ad} & B_{\mathbf{k}}^{dd}\\
B_{\mathbf{k}}^{aa} & A_{\boldsymbol{k}}^{ad\ast} & { A}_{\mathbf{k}}^{aa} & B_{\boldsymbol{k}}^{ad}\\
{A_{\boldsymbol{k}}^{ad}} & B_{\mathbf{k}}^{dd\ast} & B_{\boldsymbol{k}}^{ad\ast} & {A}_{\mathbf{k}}^{dd}
  \end{array}\right),
  \label{ddk}
\end{equation}
where the matrix elements are defined as
\be
A_{\boldsymbol{k}}^{ij}	=	-S\frac{\mu_{0}(g\mu_{B})^{2}}{2}\sum_{|{\bs r}^{ij}_{mnl}|\ne 0}\frac{|{\bs r}_{mnl}^{ij}|^{2}-3(z_{mnl}^{ij})^{2}}{|{\bs r}_{mnl}^{ij}|^{5}}e^{-i\bs{k}\cdot\boldsymbol{r}_{mnl}^{ij}},
\ee
\be
  B_{\boldsymbol{k}}^{ij}	=	-3S\frac{\mu_{0}(g\mu_{B})^{2}}{2}\sum_{|{\bs r}^{ij}_{mnl}|\ne 0}\frac{(x_{mnl}^{ij}-i y_{mnl}^{ij})^{2}}{(r_{mnl}^{ij})^{5}}e^{i\bs{k}\cdot\boldsymbol{r}_{mnl}^{ij}}.
\ee
with the superscript $i,j=a,d$ and $\boldsymbol{r}_{mnl}^{aa}=\boldsymbol{r}_{mnl}^{dd}=m{\boldsymbol v}_1+n{\boldsymbol v}_2+l{\boldsymbol v}_3$ and $\boldsymbol{r}_{mnl}^{ad}=\boldsymbol{r}_{mnl}^{aa}+(a_0,0,0)$. For a simple lattice, one can take the approximation $A_{\boldsymbol{k}}^{ad}\approx \gamma_{\bs k}A_{\boldsymbol{k}}^{aa}$ and $B_{\boldsymbol{k}}^{ad}\approx \gamma_{\bs k}B_{\boldsymbol{k}}^{aa}$ in the long wavelength regime, the dipolar Hamiltonian (\ref{ddk}) thus reduces to Eq.~(8) in Ref.~\cite{Shen20}.

The Hamiltonian due to the exchange and anisotropy reads~\cite{Shen20}
\be
H_{\boldsymbol k,-\bs k}^{0}  =\left(\begin{array}{cccc}
{\cal A} & 0 & 0 & {\cal B}_{\boldsymbol{k}}\\
0 & {\cal A} & {\cal B}_{\boldsymbol{k}} & 0\\
0 & {\cal B}_{\boldsymbol{k}} & {\cal A} & 0\\
{\cal B}_{\boldsymbol{k}} & 0 & 0 & {\cal A}
\end{array}\right).
\label{Hh0}
\ee
Here, ${\cal A}/\hbar=\omega_{\rm ex}+\omega_{\rm an}$ and ${\cal B}_{\boldsymbol k}/\hbar=\omega_{\rm ex}\gamma_{\boldsymbol k}$. The form factor $\gamma_{\boldsymbol k}=[\cos(k_x a_0)+\cos(k_y a_0)+\cos(k_z a_0)]/3$. The Hamiltonian (\ref{Hh0}) gives degenerate magnon dispersion
\be
\omega_{\bs k}=\sqrt{\omega_{\rm an}(\omega_{\rm an}+2\omega_{\rm ex})+(1-\gamma_{\bs k}^2)\omega_{\rm ex}^2}.
\ee

By combining the two Hamiltonians (\ref{ddk}) and (\ref{Hh0}) together, we calculate the bulk magnon spectrum. The dipolar-induced splitting due to the dipolar interaction in the $(k_x,k_y)$ plane is plotted in Fig.~\ref{Splitting}, which shows two gapless points at $(0,0)$ and $(\pi/a_0,\pi/a_0)$. The appearance and the locations of these gapless points are independent of the value of $k_z$, meaning that in three dimensions $(0,0,k_z)$ and $(\pi/a_0,\pi/a_0,k_z)$ are two nodal lines.

\section{Winding number for one-dimensional subsystem along $k_z$.} \label{App_B}
For the real space termination at $z$ direction, the chirality of surface modes $W(k_x, k_y)$ is always zero, suggesting its trivial topology. This can be understood as follows. For each surface, $W(k_x, k_y)=-W(-k_x, -k_y)$ due to $\{{\boldsymbol T}, {\boldsymbol C}\}=0$. In addition, both the mirror symmetry about $y$-$z$ plane ${\boldsymbol M}_{yz}$ and about $x$-$z$ plane ${\boldsymbol M}_{xz}$ anticommute with the chiral operator with ${\boldsymbol M}_{yz}H^{\rm eff}(z, k_x, k_y){\boldsymbol M}_{yz}^{-1}=H^{\rm eff}(z, -k_x, k_y)$ and ${\boldsymbol M}_{xz}H^{\rm eff}(z, k_x, k_y){\boldsymbol M}_{xz}^{-1}=-H^{\rm eff}(z, k_x, -k_y)$. This leads to $W(k_x, k_y)=-W(-k_x, k_y)$ and $W(k_x, k_y)=-W(k_x, -k_y)$. As a result, $W(k_x, k_y)\equiv 0$.

\section{Qualitative comparison of the spin wave spectra}\label{App_C}

\begin{figure}[htb]
  \includegraphics[width=8.5cm]{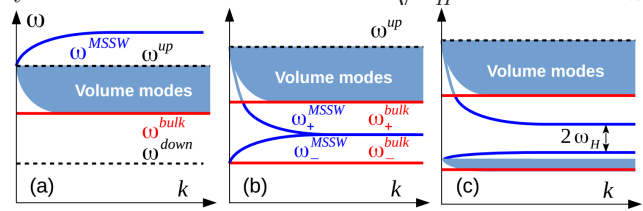}
  \caption{Qualitative comparison of the dipolar spin wave spectra in thin film of (a) FMs, (b) AFMs, and (c) AFMs with external magnetic field along the uniaxial easy axis. The red curves represent the frequencies in bulk systems, while the blue curves stands for the MSSWs in thin film. The distribution of the volume modes are illustrated by the blue shadows.}
  \label{Schematics}
\end{figure}

Figure~\ref{Schematics} illustrates qualitatively the differences between the spin wave spectra in FMs and uniaxial easy-axis AFMs, for a finite angle between wave vector $\bs k$ and magnetization direction at the equilibrium. 

For FMs, the frequency of the bulk mode represented by the red line reads
$\omega^{\rm bulk}=\sqrt{\omega_{H}^{2}+\omega_{H}\omega_{M}\sin^2{\theta_{\bs k}}}$,
whose values at $\theta_{\bs k}=0$ and $\pi/2$ give $\omega^{\rm down}=\omega_{H}$ and $\omega^{\rm up}=\sqrt{\omega_{H}^{2}+\omega_{H}\omega_{M}}$, respectively.

For AFMs, the frequencies of the two bulk modes in Eq.~(\ref{bulkmode}) gives a constant value for the lower mode $\omega_{-}^{\rm bulk}=\sqrt{\omega_{\rm an}(\omega_{\rm an}+2\omega_{\rm ex})}$ and  the upper bound of the other mode $\omega^{\rm up}=\sqrt{\omega_{\rm an}(\omega_{\rm an}+2\omega_{\rm ex}+2\omega_m)}$ in the absence of the external magnetic field. 


\bibliographystyle{prsty}

\bibliography{Refs}

\end{document}